\documentstyle[12pt]{article}
\input tcilatex

\begin{document}

\title{{\bf Leptogenesis without CP Violation \\ at Low Energies}}
\author{M. N. Rebelo
\thanks{E-mail address : mrebelo@thwgs.cern.ch
and rebelo@alfa.ist.utl.pt.
On leave from: GFP-Grupo de F\' \i sica de Part\' \i culas,
Dept de F\' \i sica,  Instituto Superior T\'ecnico,
Av. Rovisco Pais, 1049-001 Lisboa, Portugal} \\
{\it Theory Division, CERN, CH-1211 Geneva 23, Switzerland}}
\date{}
\maketitle

\begin{abstract}
In this letter we give a class of examples where
the decays of the heavy Majorana neutrinos
may violate CP even if there is no CP
violation at low energies, i.e. where
leptogenesis can take place without
Majorana- or Dirac-type CP phases
at low energies.

\end{abstract}

\setlength{\baselineskip}{14pt}
\renewcommand{\thesubsection}{\arabic{subsection}}
\begin{picture}(0,0)
       \put(335,300){CERN-TH/2002-164}
\end{picture}
\vspace{-24pt} \thispagestyle{empty}



\newpage

There is strong experimental evidence for neutrino oscillations
\cite{SK}, \cite{SNO}, thus implying non-zero neutrino masses and
mixing in the leptonic sector, together with the possibility of
CP violation. The most straightforward way of extending the
Standard Model (SM) in order to incorporate neutrino masses is
to add one neutrino field per generation, singlet under the
$SU(3)_c \times SU(2) \times U(1)$ gauge symmetry,
in analogy with the quark sector. The fact that neutrinos are
neutral particles allows for the introduction of a Majorana
mass term for the right-handed gauge singlets together with
the usual Dirac mass term, provided that lepton-number
conservation is not imposed. This leads to the see-saw mechanism
\cite{see}, which accounts in an elegant and simple way for the
smallness of neutrino masses. Leptonic CP violation may
play a crucial r\^ ole in the generation of the observed
baryon number asymmetry of the universe (BAU) via leptogenesis.
In this framework a CP asymmetry is generated through
out-of-equilibrium L-violating decays of heavy Majorana
neutrinos \cite{Fukugita:1986hr} leading to a lepton asymmetry
$L \neq 0$ while $B = 0$ is still maintained. Subsequently,
sphaleron processes \cite{Kuzmin:1985mm}, which are
$(B+L)$-violating and $(B-L)$-conserving, restore $(B+L)=0$,
thus creating a non-vanishing $B$.
Several groups have analysed the requirements on models
leading to a viable leptogenesis \cite{sev}.
At low energies the decoupling limit is an excellent
approximation and there are three CP-violating phases in
the corresponding mixing matrix, one of Dirac type,
which could be observed in neutrino oscillations \cite{CP},
and two of Majorana type, which can be interpreted in terms
of unitary triangles \cite{utri}. The question of whether or not
it is possible to establish a connexion between leptogenesis
and CP violation at low energies is very interesting and has been
addressed by several authors \cite{many}, \cite{more}.
It has been shown that, although in general this connexion
cannot be established, there are several frameworks
where the sign and size of the observed baryon asymmetry
obtained through leptogenesis can be related to CP violation
at low energies.

It is well known that in the case of three generations with no
lef-thanded Majorana mass term there are six CP-violating
phases in the leptonic sector \cite{Endoh:2000hc}.
It is possible to choose a
Weak Basis (WB) where all of these phases only
appear in the Dirac-type neutrino mass matrix.
These phases may be parametrized in such a way that
the three low-energy CP-violating phases
are a function of all of them whilst leptogenesis
can be written in terms of only three phases \cite{more}.

In this work we want to emphasize that leptogenesis can
take place even if there is no CP violation at low energies.
The prospects of finding CP-violating effects
at low energies, for instance in
future neutrino factories, are extremely exciting;
yet it is important to notice that leptogenesis
remains in principle
a viable scenario even if no CP violation
is seen at low energies.

{\bf Framework: }After spontaneous symmetry breaking, the leptonic
mass term for a minimal extension of the SM, which consists of
adding to the standard spectrum one right-handed neutrino
per generation, can be written as:
\begin{eqnarray}
{\cal L}_m  &=& -\left[ \overline{{\nu}_{L}^0} m \nu_{R}^0 +
\frac{1}{2} \nu_{R}^{0T} C M \nu_{R}^0+
\overline{l_L^0} m_l l_R^0 \right] +
{\rm h. c.} = \nonumber \\
&=& - \left[ \frac{1}{2}  n_{L}^{T} C {\cal M}^* n_L +
\overline{l_L^0} m_l l_R^0 \right] + {\rm h. c.}
\label{lm}
\end{eqnarray}
where $m$, $M$ and $m_l$ denote the neutrino Dirac mass matrix,
the right-handed neutrino Majorana mass matrix and the charged
lepton mass matrix, respectively, and
$n_L = ({\nu}_{L}^0, {(\nu_R^0)}^c)$. In this minimal
extension of the SM a term of the form
$\frac{1}{2} \nu_{L}^{0T} C m_L \nu_{L}^0$
does not appear in the Lagrangian and the matrix ${\cal M}$
is given by:
\begin{equation}
{\cal M}= \left(\begin{array}{cc}
0 & m \\
m^T & M \end{array}\right) \label{calm}
\end{equation}
with a zero entry on the (11) block. The right-handed
Majorana mass term is $SU(2) \times U(1)$
invariant; consequently it can have a value much above the
scale $v$ of the electroweak symmetry breaking, thus leading
to the see-saw mechanism.
The neutrino mass matrix $\cal M$
is diagonalized by the transformation:
\begin{equation}
V^T {\cal M}^* V = \cal D , \label{dgm}
\end{equation}
where ${\cal D} ={\rm diag} (m_{\nu_1}, m_{\nu_2}, m_{\nu_3},
M_{\nu_1}, M_{\nu_2}, M_{\nu_3})$,
with $m_{\nu_i}$ and $M_{\nu_i}$ denoting the physical
masses of the light and heavy Majorana neutrinos, respectively. It is
convenient to write $V$ and $\cal D$ in the following form:
\begin{eqnarray}
V= \left (\begin{array}{cc}
K & R \\
S & T \end{array}\right) ; \ \ \
{\cal D}=\left(\begin{array}{cc}
d & 0 \\
0 & D \end{array}\right) .
\end{eqnarray}
From Eq. (\ref{dgm}) one obtains, to an excellent approximation:
\begin{equation}
-K^\dagger m \frac{1}{M} m^T K^* =d, \label{14}
\end{equation}
together with the following exact relation:
\begin{equation}
R=m T^* D^{-1}. \label{exa}
\end{equation}
In the WB where the right-handed Majorana neutrino mass
is diagonal, it also follows to an excellent approximation that:
\begin{equation}
R=m D^{-1} .  \label{app}
\end{equation}
Equation (\ref{14}) is the usual see-saw formula with $K$
a unitary matrix.
The neutrino weak-eigenstates are related to the mass eigenstates by:
\begin{equation}
{\nu^0_i}_L= V_{i \alpha} {\nu_{\alpha}}_L=(K, R)
\left(\begin{array}{c}
{\nu_i}_L  \\
{N_i}_L \end{array} \right) \quad \left(\begin{array}{c} i=1,2,3 \\
\alpha=1,2,...6 \end{array} \right) ,
\label{15}
\end{equation}
and thus the leptonic charged-current interactions are given by:
\begin{equation}
- \frac{g}{\sqrt{2}} \left( \overline{l_{iL}} \gamma_{\mu} K_{ij}
{\nu_j}_L +
\overline{l_{iL}} \gamma_{\mu} R_{ij} {N_j}_L \right) W^{\mu}
+{\rm h.c.}
\label{16}
\end{equation}
From Eqs. (\ref{15}), (\ref{16}) it follows that $K$ and $R$ give the
charged-current couplings of charged leptons to the light
neutrinos $\nu_j$ and to the heavy
neutrinos $N_j$, respectively. The unitary
matrix $K$, which contains all
the information about CP violation at low energies,
can be parametrized as:
\begin{equation}
K= P_{\xi} {\hat U_{\rho}} P_{\alpha}
\ \ \ \longrightarrow \ \ \ {\hat U_{\rho}} P_{\alpha}
\label{kkk}
\end{equation}
with $P_{\xi}={\rm diag}\left(\exp(i\xi_1),\exp(i\xi_2),\exp(i\xi_3)
\right)$, and
$P_\alpha ={\rm diag}(1, \exp(i\alpha_1) \exp(i\alpha_2))$
leaving ${\hat U_{\rho}}$ with only one phase as in the case of the
Cabibbo, Kobayashi and Maskawa matrix.
Since $ P_{\xi}$ can be rotated away
by a redefinition of the charged leptonic fields, $K$ is left
with three CP-violating phases, one of Dirac type $\rho$ and two
of Majorana character $\alpha_1$ and $\alpha_2$.

The computation of the lepton-number asymmetry, in this extension
of the SM, resulting from the decay of a heavy Majorana neutrino $N^j$
into charged leptons $l_i^\pm$ ($i$ = e, $\mu$ , $\tau$) leads to
\cite{sym} :
\begin{eqnarray}
A^j &=& \frac{g^2}{{M_W}^2} \sum_{k \ne j} \left[
{\rm Im} \left((m^\dagger m)_{jk} (m^\dagger m)_{jk} \right)
\frac{1}{16 \pi} \left(I(x_k)+ \frac{\sqrt{x_k}}{1-x_k} \right)
\right]
\frac{1}{(m^\dagger m)_{jj}},   \nonumber \\
&=& \frac{g^2}{{M_W}^2} \sum_{k \ne j} \left[ (M_k)^2
{\rm Im} \left((R^\dagger R)_{jk} (R^\dagger R)_{jk} \right)
\frac{1}{16 \pi} \left(I(x_k)+ \frac{\sqrt{x_k}}{1-x_k} \right)
\right]
\frac{1}{(R^\dagger R)_{jj}} \nonumber \\
\label{rmy}
\end{eqnarray}
with the lepton-number asymmetry from the $j$ heavy Majorana
particle, $A^j$, defined in terms of the family number asymmetry
$\Delta {A^j}_i={N^j}_i-{{\overline{N}}^j}_i$ by :
\begin{equation}
A^j = \frac{\sum_i \Delta {A^j}_i}{\sum_i \left({N^j}_i +
\overline{N^j}_i \right)}
\label{jad}
\end{equation}
the sum in $i$ runs over the three flavours
$i$ = e $\mu$ $\tau$, $M_k$ are the heavy neutrino masses,
the variable $x_k$
is defined as  $x_k=\frac{{M_k}^2}{{M_j}^2}$ and
$ I(x_k)=\sqrt{x_k} \left(1+(1+x_k) \log(\frac{x_k}{1+x_k}) \right)$.
From Eq.~(\ref{rmy}) it can be seen that the lepton-number
asymmetry is only sensitive to the CP-violating phases
appearing in $m^\dagger m$ in the WB, where $M$ and $m_l$
are diagonal (or equivalently in $R^\dagger R$).

{\bf Leptogenesis with no CP violation at low energies: } Let us
go to the WB where $M$ and $m_l$ are diagonal, real and positive
($M \equiv D $) and choose a matrix $m$ of the form
\cite{param}:
\begin{equation}
m= i {\hat U_{\rho}} P_{\alpha} {\sqrt d} O^c {\sqrt D },
\label{mmm}
\end{equation}
where ${\sqrt d}$ and ${\sqrt D }$ are diagonal real
matrices such that   ${\sqrt d} {\sqrt d} = d $,
${\sqrt D } {\sqrt D } = D $ and $O^c$
is an orthogonal complex matrix, i.e.  $O^c {O^c}^T = 1 $
but $O^c {O^c}^\dagger \neq 1 $. In this WB all CP-violating
phases appear in $m$. From Eq.~(\ref{mmm}) together
with  Eq.~(\ref{14}) we
obtain the matrix $K$ given by Eq.~(\ref{kkk})
and in general it will
violate CP. The physical relevance of this expression for
the study of viable leptogenesis and its connection with
low energy physics was emphasized by I. Masina at SUSY02
\cite{masina}.
Particularizing for $\alpha_1 = \alpha_2 = 0 $ together with
$\rho = 0$, there is no CP violation at low energies. Yet
leptogenesis is sensitive to
the combination $m^\dagger m$,  which is given by:
\begin{equation}
h \equiv m^\dagger m = {\sqrt D }  {O^c}^\dagger
d   O^c {\sqrt D };
\label{hhh}
\end{equation}
consequently, provided that the combination
${O^c}^\dagger  d   O^c $ is CP-violating, we may have
leptogenesis even without CP violation at low energies either
of Dirac or Majorana type.

It is possible to write WB-invariant conditions which have to
vanish in order for CP invariance to hold. The non-vanishing of
any of these invariants signals CP violation
\cite{Branco:gr}. In Ref. \cite{more} the following
WB invariants, sensitive to CP-violating phases
that appear in leptogenesis, were derived:
\begin{eqnarray}
I_1  \equiv {\rm Im Tr}[h H M^* h^* M]=  \nonumber \\
= M_1M_2(M_2^2 - M_1^2) {\rm Im} (h_{12}^2)
+ M_1M_3(M_3^2 - M_1^2) {\rm Im} (h_{13}^2)+ \nonumber \\
+ M_2M_3(M_3^2 - M_2^2) {\rm Im} (h_{23}^2);
\nonumber
\label{i1} \\
I_2 \equiv {\rm Im Tr}[h H^2 M^* h^* M]= \nonumber \\
= M_1M_2(M_2^4 - M_1^4) {\rm Im} (h_{12}^2)
+ M_1M_3(M_3^4 - M_1^4) {\rm Im} (h_{13}^2)+ \nonumber \\
+ M_2M_3(M_3^4 - M_2^4) {\rm Im} (h_{23}^2);
\nonumber
\label {i2l} \\
I_3 \equiv {\rm Im Tr}[h H^2 M^* h^* M H]\nonumber \\
= M_1^3 M_2^3 (M_2^2 - M_1^2) {\rm Im} (h_{12}^2)
+ M_1^3 M_3^3 (M_3^2 - M_1^2) {\rm Im} (h_{13}^2)+ \nonumber \\
+ M_2^3 M_3^3 (M_3^2 - M_2^2) {\rm Im} (h_{23}^2)
\label{i3l}
\end{eqnarray}
The second equality for each $I_i$ corresponds to the evaluation
of these WB invariants in the WB where the right-handed neutrino
mass is diagonal, with $M_i$ the corresponding diagonal elements.
The matrix $H$ is defined by $H \equiv M^{\dagger} M$.

Choosing the matrix $O^c $ of the form:
\begin{equation}
O^c = A_{12} \cdot  A_{23} \cdot  A_{13}
\end{equation}
with
\begin{eqnarray}
A_{12}\ =\ \left(
\begin{array}{ccc}
\cosh \theta_{12} & i \sinh \theta_{12} & 0 \\
-i \sinh \theta_{12} &\cosh \theta_{12}  & 0 \\
0 & 0 & 1
\end{array}
\right), \  A_{13}\ =\ \left(
\begin{array}{ccc}
\cosh \theta_{13} & 0 & i \sinh \theta_{13}  \\
0 & 1 & 0 \\
-i \sinh \theta_{13} & 0 & \cosh \theta_{13}
\end{array}
\right), \nonumber  \\
A_{23}\ =\ \left(
\begin{array}{ccc}
1 & 0 & 0 \\
0 &\cosh \theta_{23}  & i \sinh \theta_{23}  \\
0 & -i \sinh \theta_{23} &  \cosh \theta_{23}
\end{array}
\right),
\label{eq6}
\end{eqnarray}
we obtain (simplifying the notation with cosh and sinh
replaced by ch and sh):
\begin{eqnarray}
{\rm Re} h_{12} &=& ({\rm sh}^2 \theta_{12} {\rm sh} \theta_{13}
{\rm sh}  \theta_{23} {\rm ch}  \theta_{23} d_1 +
{\rm sh}  \theta_{13} {\rm sh}  \theta_{23} {\rm ch}^2  \theta_{12}
{\rm ch}  \theta_{23} d_2 + \nonumber \\
&+& {\rm sh}  \theta_{13} {\rm sh}  \theta_{23} {\rm ch}  \theta_{23}
d_3 ) \sqrt M_1 \sqrt M_2 ; \nonumber \\
{\rm Im} h_{12}  &=& i {\rm sh} \theta_{12} {\rm ch} \theta_{12}
{\rm ch}  \theta_{13} {\rm ch}  \theta_{23} (d_1 + d_2)
\sqrt M_1 \sqrt M_2 ; \nonumber \\
{\rm Re} h_{13} &=& - {\rm sh} \theta_{12} {\rm sh} \theta_{23}
{\rm ch}  \theta_{12}  (d_1 + d_2)
\sqrt M_1 \sqrt M_3 ; \nonumber \\
{\rm Im} h_{13} &=& i [({\rm sh} \theta_{13} {\rm ch}^2 \theta_{12}
{\rm ch}  \theta_{13} + {\rm sh}^2  \theta_{12}
{\rm sh} \theta_{13} {\rm sh}^2 \theta_{23} {\rm ch} \theta_{13}) d_1 +
\nonumber \\
&+& ({\rm sh}^2 \theta_{12} {\rm sh} \theta_{13}
{\rm ch}  \theta_{13} + {\rm sh}  \theta_{13}
{\rm sh}^2 \theta_{23} {\rm ch}^2 \theta_{12} {\rm ch} \theta_{13}) d_2 +
\nonumber \\
&+& {\rm sh} \theta_{13} {\rm ch} \theta_{13} {\rm ch}^2  \theta_{23}
d_3 ] \sqrt M_1 \sqrt M_3 ; \nonumber \\
{\rm Re} h_{23} &=&  {\rm sh} \theta_{12} {\rm sh} \theta_{13}
{\rm ch}  \theta_{12} {\rm ch}  \theta_{23} (d_1 + d_2)
\sqrt M_2 \sqrt M_3 ; \nonumber \\
{\rm Im} h_{23}  &=& i ({\rm sh}^2 \theta_{12} {\rm sh} \theta_{23}
{\rm ch} \theta_{13} {\rm ch}  \theta_{23} d_1 +
{\rm sh} \theta_{23}
{\rm ch}^2  \theta_{12} {\rm ch} \theta_{13}  {\rm ch} \theta_{23} d_2
\nonumber \\
&+& {\rm sh} \theta_{23} {\rm ch} \theta_{13} {\rm ch}  \theta_{23} d_3)
\sqrt M_2 \sqrt M_3. \label{iii}
\end{eqnarray}
The $d_i$ are the diagonal elements of the matrix $d$
(the masses of the light neutrinos).
In general ${\rm Im} h_{ij}^2, i\neq j $ do not vanish
and there is leptogenesis. On the other hand, if any
of the $\theta _{ij}$ is zero, these imaginary parts vanish
since all products ${\rm Re} h_{ij}
{\rm Im} h_{ij}$ contain the factor
${\rm sh} \theta_{12} {\rm sh} \theta_{13} {\rm sh} \theta_{23}$.
Equations (\ref{i3l}) are no longer useful to discuss CP
violation in the limit $M_1 = M_2 = M_3$ since in this case
they vanish trivially, although this degeneracy
does not necessarily imply CP conservation at high energies
\cite{more}.

{\bf Final Additional Comments: }
In this framework
low-energy physics only enters Eq.~(\ref{hhh})
through the masses of the light neutrinos, which are
already constrained by experiment. In fact,
Eq.~(\ref{hhh}) has no explicit dependence on mixing and
CP violation at low energies, since $K$ cancels out.
With the present experimental
knowledge there is freedom in the choice of
the masses of the heavy neutrinos. Furthermore,
low-energy physics is insensitive to the
matrix $O^c$. As a result one can only establish a connection
between leptogenesis and CP violation at low energies
in models where additional constraints are imposed,
so that, for instance,
the matrix $D$ is no longer independent of
$K$.

{\bf Acknowledgements}: The author thanks G. C. Branco for useful
comments and reading the manuscript and
the Theory Division of CERN for hospitality.
This work was partially supported by
FCT (``Funda\c c\~ao
para a Ci\^encia e a Tecnologia'', Portugal) through projects
CERN/FIS/43793/2001 and CFIF - Plurianual (2/91).

\end{document}